# A Quantum Walk Comb Source at Telecommunication Wavelengths

**Bahareh Marzban**[1], **Lucius Miller**[1], **Alexander Dikopoltsev**[1], **Mathieu Bertrand**[1], **Giacomo Scalari**[1], **Jérôme Faist**[1]

[1]Institute of Quantum Electronics, ETH Zürich, 8093 Zürich, Switzerland, and Quantum Center, ETH Zürich, 8093 Zürich, Switzerland
*Corresponding authors: bmarzban@ethz.ch, jfaist@ethz.ch

**Abstract:**
We demonstrate a quantum walk comb in synthetic frequency space formed by externally modulating a semiconductor optical amplifier operating in the telecommunication wavelength range in a unidirectional ring cavity. The ultrafast gain saturation dynamics of the gain medium and its operation at high current injections is responsible for the stabilization of the comb in a broad frequency modulated state. Our device produces a nearly flat broadband comb with a tunable repetition frequency reaching a bandwidth of 1.8THz at the fundamental repetition rate of 1GHz while remaining fully locked to the RF drive. Comb operation at harmonics of the repetition rate up to 14.1GHz is also demonstrated. This approach paves the way for next-generation optical frequency comb devices with potential applications in precision ranging and high-speed communications.

**Introduction:**
Optical frequency combs are discrete, evenly spaced single frequency waveforms that are precisely phase-locked to one another[1]. They are unique tools for both scientific research and technological applications[2]. By leveraging advanced lithography and nanofabrication techniques[3], integrated optical frequency comb sources drastically reduce system size, weight, power consumption and cost[4]. As a result, they are not only highly effective in applications like optical spectroscopy[5,6], remote sensing[7], and broadband wavelength-division multiplexing[8,9] but also hold great promise for a broader range of consumer technologies, including autonomous driving[10], 5G/6G communications[11], and machine learning[12].

In the realm of integrated photonic platforms, significant advancements have been made to miniaturize the laser cavity and the nonlinear frequency-generation component. Frequency comb sources, such as Kerr[13], electro-optical[14], and quadratic combs[15], rely on external pump lasers injected into a nonlinear medium generating a comb via optical non-linearities. In contrast, semiconductor mode-locked lasers produce the comb directly within the active medium, achieving relatively high wall plug efficiencies while providing key advantages in terms of operational simplicity[16,17]. One locking mechanism used in semiconductor mode locked lasers is active mode-locking, where the loss of a section of the laser cavity is resonantly modulated, opening a gain window in the time domain for the propagation of a pulse.

The problem of resonant modulation of a cavity at the roundtrip frequency can be studied by treating the successive longitudinal modes of the laser resonator as lattice sites of a synthetic dimension[18]. In this picture, the nearest-neighbor coupling caused by the modulation at the round-trip frequency leads to quantum-walk dynamics[18–20], where the photons coherently diffuse in the lattice[21]. In an early application of this approach, the author in[22] shows that the equation that describes active mode-locking in the frequency domain, which includes dispersion and gain curvature, is mathematically analogous to the quantum harmonic oscillator with Hermite-Gauss solutions in the limit of vanishing mode spacing. As shown schematically in Fig. 1a, each solution corresponds to a different amplitude and phase combination of the individual longitudinal modes. In media with slow gain recovery—where the gain recovery time ($T_1$) greatly exceeds the cavity round-trip time ($T_{rep}$), as in solid-state laser systems—the dissipative mechanism stabilizes the system in its lowest-energy state: a Gaussian pulse with a bandwidth scaling as the fourth root of the modulation depth. The author of [22] showed that higher-order, broader frequency solutions are inherently unstable in this regime.

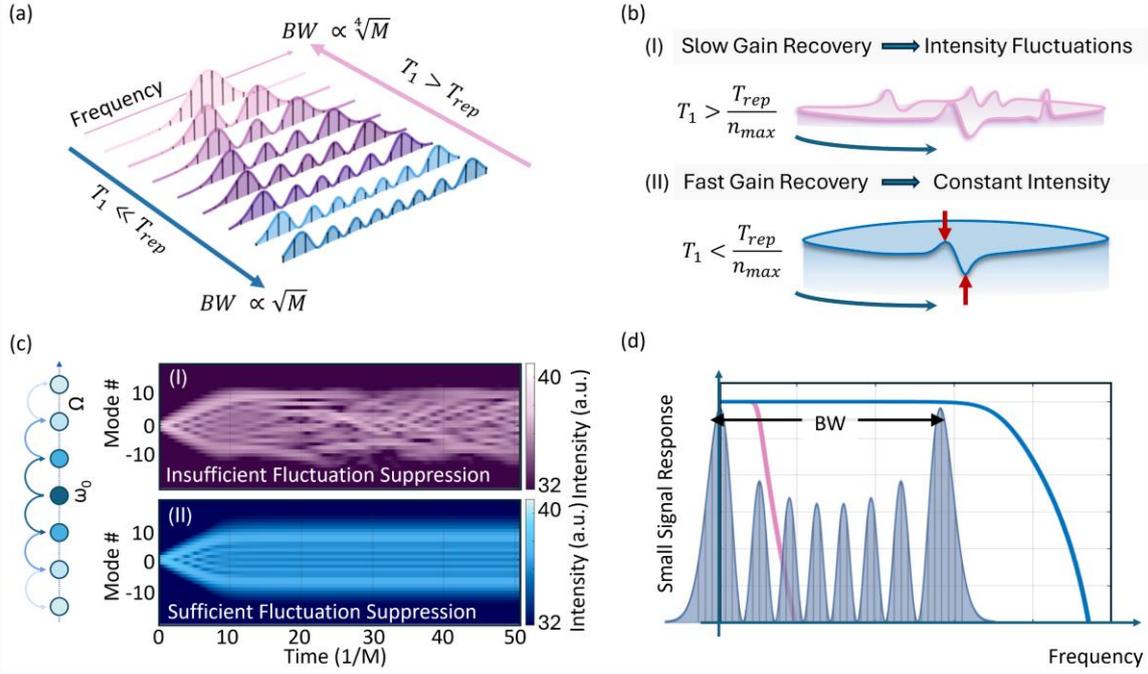

**Figure 1| A quantum Walk Comb Laser. a,** Upon RF injection, the comb laser's operational state depends on the relationship between the carrier recovery $T_1$ and $T_{rep}$. If $T_1 \gg T_{rep}$ the laser stabilizes on the ground state and pulses, whereas if $T_1 \ll T_{rep}$ the laser stabilizes on a broad super mode with its bandwidth limited only by dispersion and gain bandwidth. **b,** Under resonant modulation, (I) a laser with slow gain recovery cannot suppress intensity fluctuations, leading to amplitude modulation. In contrast, (II) a laser with fast gain recovery effectively suppresses these fluctuations, resulting in a comb with a flat-top spectrum and constant intensity. **c,** Left: A RF tone at a frequency matching the cavity mode spacing is superimposed on the drive current of a single-mode laser, inducing coupling between adjacent modes that can be seen as lattice sites in a synthetic dimension. Right: (I) When the gain recovers slower than the fastest fluctuation of the waveform, the laser cannot suppress the intensity fluctuations and hence it does not stabilize. (II) When gain recovery is faster than the fastest fluctuations, upon resonant modulation the laser spectrum broadens and stabilizes on a broad super mode. **d,** By increasing the bias current far above threshold, the intracavity light intensity of an interband semiconductor diode laser is increased, switching it to a fast gain recovery state that stabilizes the comb into a broad, flat-top spectrum. In this case, the gain recovers fast enough to suppress intensity fluctuations of the largest frequency components in the spectrum.

In contrast, it was recently shown[19] that these higher order solutions are stable in a system where the gain recovery time satisfies $T_1 \ll T_{rep}$. Indeed, a fast gain saturation will lower the net gain of states with large intracavity intensity fluctuations, such as the Gaussian fundamental state, and stabilize the highest excited state, exhibiting an almost constant intensity (schematically shown in Fig. 1b). Those states reach the maximal bandwidth of the system, equal in an analytical approximation to $2c/\pi n \sqrt{(M/\beta)}$ where M the modulation depth and β the GVD parameter[19]. The available bandwidth of this frequency modulated comb represents a quadratic improvement over traditional active mode locked pulsed states.

Recently, a quantum walk comb laser was experimentally demonstrated, relying on the ultrafast intersubband gain recovery time ($T_1 \sim 0.4ps$[19]) of a mid-infrared quantum cascade laser for the stabilization of a broad and flat optical frequency comb spectrum. However, using the fast saturation of intersubband transitions limits this approach to mid-infrared devices.

It was argued that because of the relatively long radiative lifetime of band-to-band transitions, interband semiconductor lasers were slow-gain devices satisfying $T_1 \gg T_{rep}$[23], precluding their use for the realization of a quantum walk comb. Here we show that this argument is an oversimplification and present a broad-bandwidth quantum walk comb laser that exploits the fast fluctuation suppression response of an interband gain material (see Fig. 1c (I) and (II)). By incorporating a multi-quantum well layer stack as a gain medium within a unidirectional ring cavity and operating the laser at high intracavity intensities, we effectively enhance the gain recovery rates, allowing the quantum walk comb to exploit the bandwidth potential of the optical gain (Fig. 1d). This approach not only significantly improves the stability of the comb lines but also provides precise control over the comb's spectral characteristics. Operating in the NIR range, we demonstrate a broad 1.8THz comb, ideal for applications such as LiDAR, and 5G/6G technologies and communications.

**Quantum Walk Comb Laser in the NIR**

In fact, the dynamical characteristics of interband gain media comprise a rich range of phenomena with very different characteristic times, as shown experimentally by pump-probe measurements using short pulse excitation[24]. Indeed, thermalization of the population after spectral hole burning is achieved by carrier-carrier scattering with characteristic times in the 100fs or below. Cooling of the electron-hole population is driven by optical phonons yielding a carrier lifetime on the order of 200fs. While the band-to-band recovery time is 1ns or longer in non-lasing operation, this time is much shorter and dominated by stimulated emission in lasing devices operated well above threshold. As a result, the lifetime can be driven down to a few picoseconds, as shown by the disappearance of the relaxation oscillation in the modulation characteristics of semiconductor lasers driven at high optical fields[25]. In this respect, semiconductor lasers, when operated well above threshold, have the dynamical properties of fast gain devices. Their self mode-locking characteristics are therefore much closer to the ones of quantum cascade lasers than of solid state lasers, as shown by the numerous reports of FM comb self mode-locking operation[26–28].

Our quantum walk comb laser consists of an external cavity ring laser, comprising of a booster optical amplifier based on a compressively strained InGaAsP/InP MQW layer stack. The optical amplifier has a 85nm gain bandwidth, a 30dB small signal gain and an 18dBm output saturation power at 600mA current injection. As shown in Fig. 2a, the light is coupled in and out of the optical amplifier via a pair of lenses while the ring cavity is formed by two mirrors, a 90/10 beam splitter for the extraction of the output beam, and an optical isolator providing 45dB isolation ensuring unidirectional lasing. As a result, spatial hole burning and the corresponding longitudinal multimode instabilities are prevented, yielding in the absence of modulation a single mode device at a wavelength close to 1600nm with a side mode suppression ratio of 35dB. The device is then operated at a current well above its lasing threshold of 85mA (see supplementary material for the light-current characteristics). A quantum walk comb is achieved by applying a modulation signal at a frequency that matches the inverse of the cavity round-trip time to the driving current of the optical amplifier, leveraging on the phase modulation induced by the linewidth enhancement factor[29], stabilizing the lasing state onto a broad supermode with a flat-top spectrum.

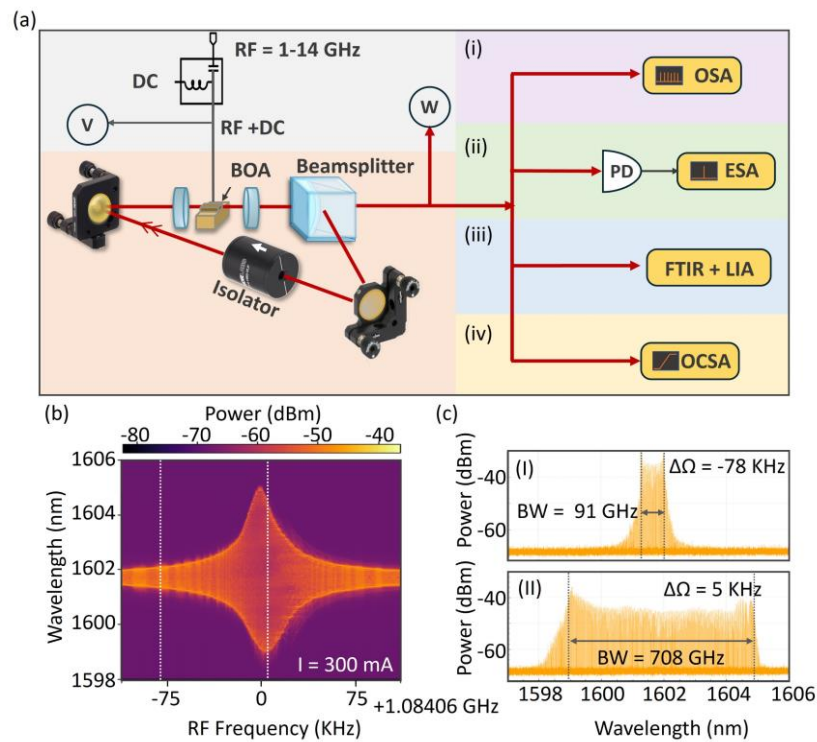

**Figure 2 |** External cavity quantum walk comb laser. **a**, Experimental setup for comb characterization, including (i) optical spectrum measurements, (ii) RF beat note measurements, (iii) spectral phase measurements, and (iv) time-resolved measurements. Components include a booster optical amplifier (BOA), two lenses, a 90/10 beam splitter, an isolator, and two dielectric mirrors. A bias-T injects DC and AC currents from an RF source into the gain chip. LIV characteristics are measured using a voltmeter (V) and power meter (W). Abbreviations: OSA (Optical Spectrum Analyzer), ESA (Electrical Spectrum Analyzer), RF (Radio Frequency), FTIR (Fourier-transform infrared spectroscopy), LIA (Lock-In Amplifier). **b**, RF frequency injection sweeps reveal that bandwidth increases as detuning from the cavity roundtrip frequency decreases. **c,d**, Spectra at RF injection frequencies ΔΩ = 78 and 5KHz where Ω is the resonance frequency, resulting in combs with 91GHz and 708GHz bandwidths, respectively.

To study the effect of the spectrum broadening as a function of RF injection frequency, we measure the steady-state spectra of our device using an optical spectrum analyzer (OSA). As shown in Fig. 2b, we measure the optical spectrum versus detuning from the cavity round trip frequency by conducting a frequency sweep, injecting a 5dBm strong signal at a frequency close to the cavity resonance at $\Omega$ = 1.08406GHz, with increments of 1kHz. As was observed in[19], when the RF modulation frequency approaches resonance, the spectrum broadens significantly and controllably, its bandwidth increasing in this case from 91 GHz at 78kHz detuning to 708GHz at resonance (Fig. 2c (I) and (II)).

Shown in Fig 3a and expected for such a comb, at small frequency detuning ($\Delta\Omega$ < 35kHz) the maximum bandwidth reached follows the expected dependence of $BW = 2c/\pi n \sqrt{(M/\beta)}$ as a function of modulation power when the latter is changed from -20dBm to 18dBm. By changing the drive current setpoint of the optical amplifier from 300mA to 500mA, with RF injected power ranging from -3 to 18dBm, a broad maximum optical bandwidth of 1.78THz is observed. However, this bandwidth slightly deviates from the analytical predictions mentioned above. We attribute this discrepancy to higher-order nonlinearities, which are not accounted for in the analytical model, becoming significant at high intracavity intensities. Additionally, uneven gain curvature at this operating point may contribute to the observed deviation.

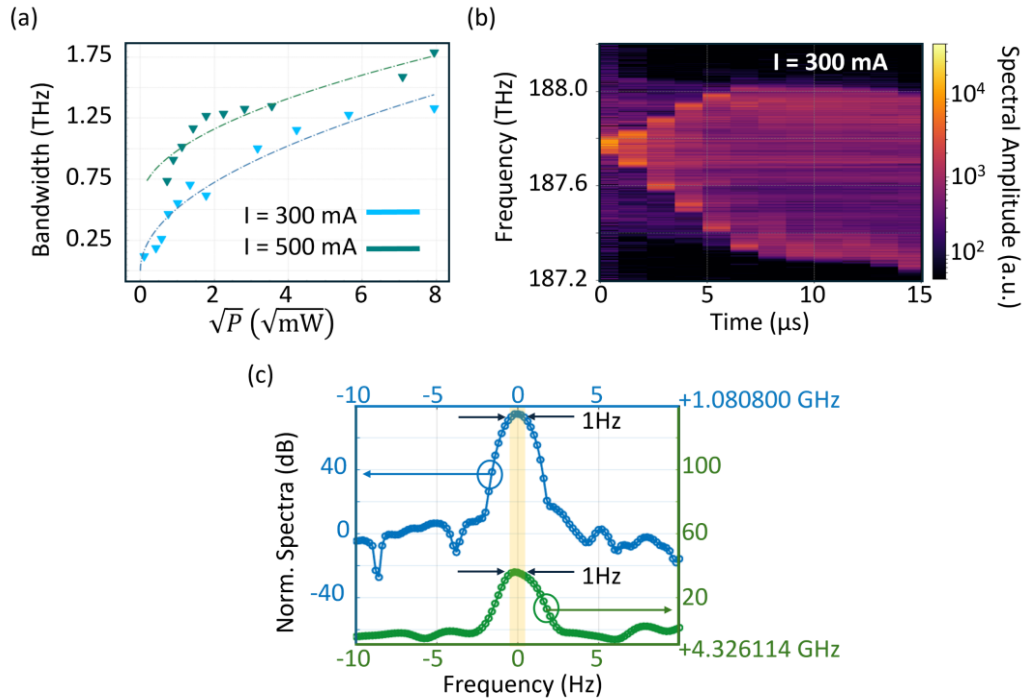

**Figure 3|** Characteristics of a quantum walk comb laser. **a,** Spectral bandwidth as a function of square root of modulation power P, showing the expected square root behavior. **b,** Time-resolved measurements reveal linear ballistic expansion of the bandwidth upon RF injection, leading to stabilization of on broad comb after approximately 8.6μs. The time-resolved data were obtained at an RF power of 10dBm. **c,** A 1Hz intermode beat note is observed across a broad 1.58THz spectrum (blue curve) when the injected frequency matches the fundamental frequency. The green curve shows the beat note between two comb lines when the fourth harmonic is injected, also exhibiting a 1Hz 3dB linewidth.

We further investigate the transition from single-mode laser operation to a broad comb state under 10dBm RF injection, confirming the laser's ballistic bandwidth expansion through time-resolved measurements using a boxcar averaging technique on a home-made Fourier-transform infrared spectrometer (see Methods). Given the relatively long cavity length as compared to chip-scale devices resulting in a round-trip time of approximately 1ns, the light requires about 8.6μs to fully expand into the final comb bandwidth. This observation confirms the linear expansion and stabilization into the broad state, as shown in Fig. 3b. From the expansion data for Fig. 3b at I = 300mA and RF injection of 10dBm we extract a value of GVD of 12780fs$^2$/mm (see supplementary).

Beyond allowing control over the laser bandwidth, this frequency comb source exhibits remarkable stability. To highlight the latter, we measure the intermodal beat note at 500mA injection current and an RF injection = 17dBm allowing a bandwidth of 1.58THz to be reached. An exceptionally narrow RF beat note of 1Hz is observed using a 1Hz resolution bandwidth on a Keysight N9040B UXA RF spectrum analyzer, as shown in Fig. 3c. This narrow linewidth is consistently observed when two individual comb lines are filtered, and the intermodal beat note is measured (green curve in Fig. 3c). As expected for our mode of operation, the RF linewidth mirrors the stability of the source. In fact, contrary to spontaneously formed frequency combs, the modulation removes the translation degeneracy of the signal in the cavity, and contributes to the reduced noise and enhances the stability of the actively mode-locked frequency comb[30].

When the RF injection exactly matches resonance, the quantum walk comb is predicted to stabilize on a frequency modulated state characterized by an almost constant intensity and an instantaneous frequency that follows a half sinusoidal function of time. To verify these properties, we measure the complex spectra of the laser near resonance with an optical complex spectrum analyzer. The amplitude and intermodal phases, i.e., the phase difference between adjacent modes of the various comb lines is recorded and shown in Fig. 4a(I) and 4b(I) for an RF injection power of P = 5 dBm and 18dBm respectively. As shown in Fig. 4 b, as we move from the low to the high frequency end of the comb, the phase difference $\Delta\phi$ between adjacent modes do sweep a full $2\pi$, indicating that the group delay of the extremes of the spectrum lie at the opposite end of the period. This behavior is consistent with the phase locking mechanism that arises from fast gain saturation, dynamically adjusting the phase to minimize intensity fluctuations within the cavity and producing a quasi-linear chirp of the instantaneous frequency as a function of time. Accordingly, as shown in Fig. 4c and d where these experimental measurements are compared to simulations as well as to an analytical solution of the quantum walk comb equation, the instantaneous frequency exhibits the expected half-sinusoid function of time followed by a sharp transition to the initial point at the end of the period.

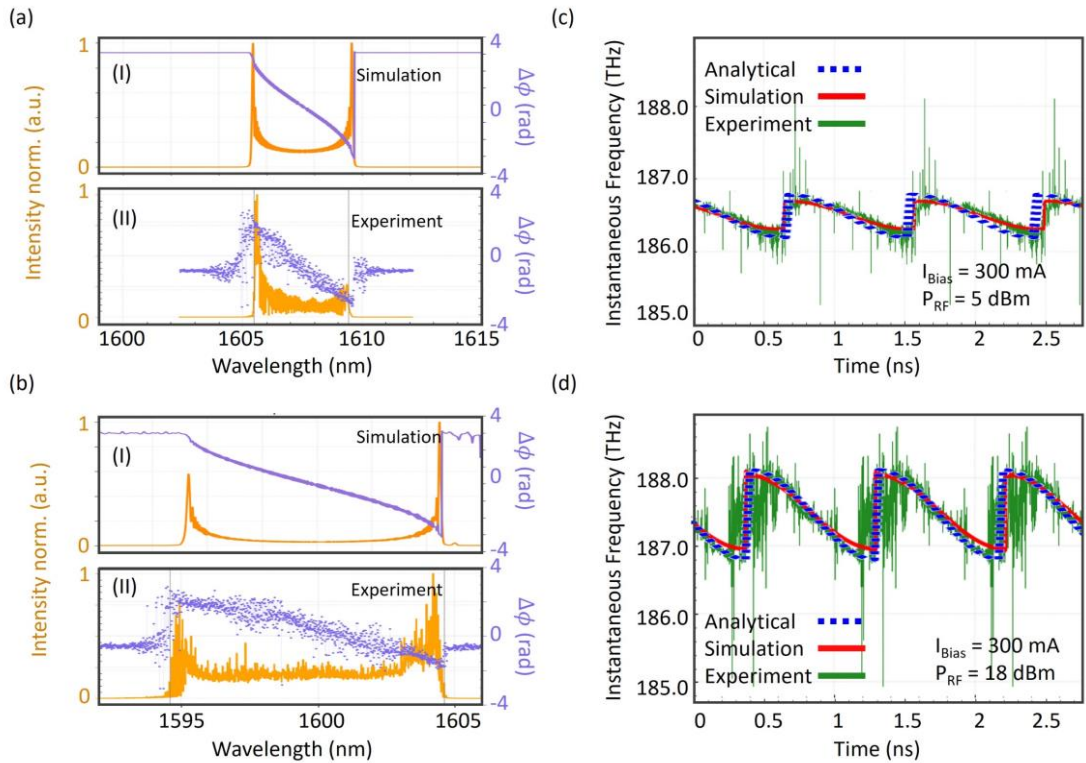

**Figure 4|** Complex optical spectra measured at RF power = 5dBm, 18dBm atop a 300mA DC current. **a, b,** Relative phase of the comb lines can be seen to span over 2π. The phase relationship is chosen by the laser to minimize intensity modulation. (I) and (II) Modeling and experimental results are presented, both showing normalized power on a linear scale. **c, d,** Instantaneous frequency measurements over 3 periods as the laser is operated near resonance. A rapid shift from minimum to maximum wavelength over very short timescales is observed. Both modeling and measurement results are illustrated.

In Fig. 4, we model the resulting comb spectral shape, which stabilizes into a broad, flat-topped state, using a modified form of the complex Ginzburg-Landau equation (CGLE)[19,31]. The system's dynamics are modeled numerically using the split-step method, with Fig.4a(II) and 4b(II) showing the resulting spectra and relative phase after 10000 round trips. Notably, the measured instantaneous frequency is in excellent agreement with both the analytical and numerically obtained results.

The laser shows a large flexibility, allowing both large bandwidths and harmonic mode spacing to be achieved. First, Fig. 5a demonstrates the 1.8THz bandwidth, notably without dispersion compensation, using 18dBm of RF power on top of a 500mA DC current. Second, akin to other actively mode-locked lasers, modulation can be applied not only at the roundtrip frequency but also higher harmonics. We investigate the effects of higher harmonic modulation, with two examples presented in Fig. 5b and 5c, with RF injection frequencies of 3×Ω and 13×Ω, respectively. Modulation at the third harmonic with 19.5dBm RF power achieved a spectrum width of 955GHz, while modulation at the thirteenth harmonic with 10dBm RF power yielded a spectrum width of 522GHz. The limited bandwidth response of the laser at 14 GHz (13×Ω) is attributed to inadequate RF termination. Compared to passive harmonic mode locked systems[32], our approach allows for dynamic adjustment of multiple free spectral ranges (FSRs) by simply varying the RF injection frequency, without the need for physical modifications to the laser. In addition to the comb lines with frequency separations of 3×Ω or 13×Ω, we also observe lines with the fundamental frequency separation of 1×Ω, that are 20-30dB weaker, which we assume are the result of weak nonlinearities in the system.

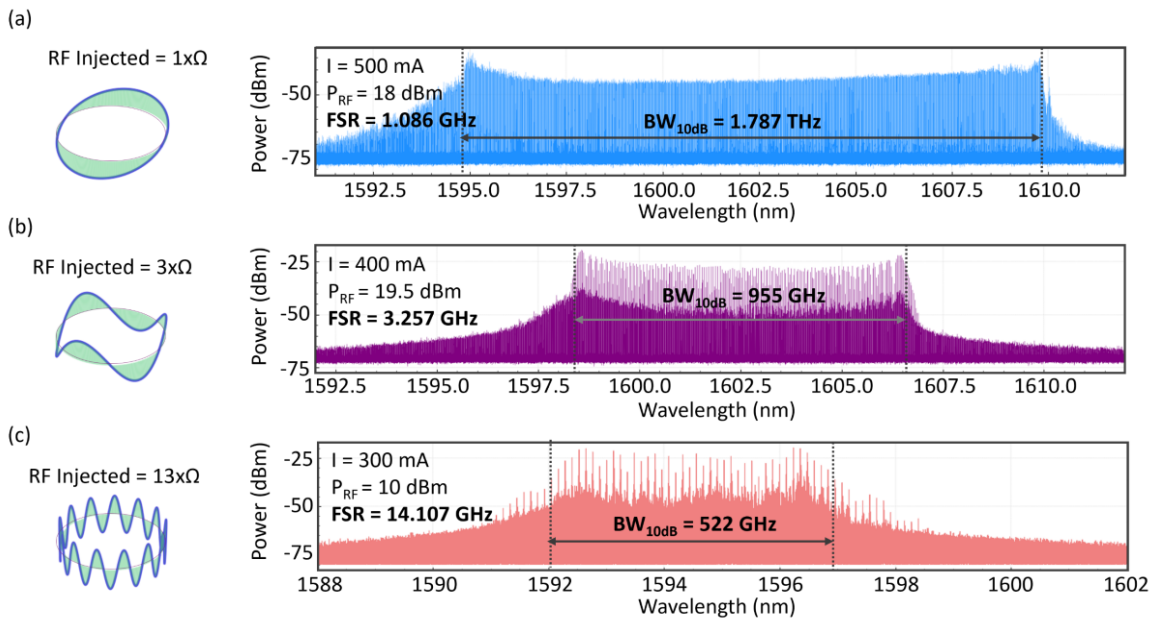

**Figure 5| Controllability of the comb laser. a,** Increasing the current to 500 mA and applying an RF power injection of 18dBm achieves a 10dB bandwidth of 1.787THz. Panels (b) and (c) illustrate two examples of harmonic injection: with 955GHz and 522GHz 10dB bandwidths for 3×Ω and 13×Ω harmonics, respectively.

**Discussion and Outlook**

We experimentally demonstrate a highly controllable mode-locked quantum walk comb laser operating in the NIR wavelength range. Contrary to the conventional view that interband lasers are inherently slow, we show that, when operated at an optimal point, these lasers exhibit fast gain recovery dynamics like those of QCLs, enabling the formation of a robust quantum walk comb.

**Methods**

**LIV Characterization**

Voltage for the LIV is measured by measuring the voltage between shield and center conductor from a tap taken just before the bias tee using a voltmeter (HP 34401A) in 10 MΩ input impedance mode. Power is measured by placing a NIR power meter (Thorlabs PD100USB, S132C Sensor) in the output beam.

**Time-resolved FTIR spectroscopy**

A Zurich Instruments UHFLI Lock-In amplifier in boxcar-averager mode is used to gate the RF-generator (Rohde-Schwarz SMB100A) using a 20kHz square wave. The signal of a fast photodetector (Newport, 818-BB-35, 15GHz 3 dB bandwidth) at the output of the FTIR spectrometer is averaged for 250 ns at a sweepable offset from the internally kept gate signal using the boxcar averager. The lock-in output holds the measured voltage until the next pulse. The average detector output, as well as a reference interferogram from a fiber-coupled laser diode (RIO GRANDE) for measurement of the stage delay are digitized using an oscilloscope (LeCroy HDO6104). By scanning the time offset of the gate from the trigger signal, the optical spectrum of the laser as a function of time relative to the onset of modulation can be recovered using conventional FTIR spectroscopy data analysis. The interferograms are cropped to two bursts and resampled using the zero-crossings of the reference interferograms.

**Acknowledgements**:
The authors would like to thank Dr. Philipp Täschler and Barbara Schneider from the Quantum Optoelectronics Group at ETH Zurich for their valuable discussions and for providing initial code related to time-resolved measurements.
**Funding**:

This work was supported by the following: MIRAQLS: Staatssekretariat für Bildung, Forschung und Innovation SBFI (22.00182) in collaboration with EU (grant Agreement 101070700); ETH Fellowship program: (22-1 FEL-46) (to A.D.)

**Author Contributions:**
B.M built the setup, with assistance from M.B. B.M performed the simulations based on predictive modeling of A.D. B.M and L.M performed the characterizations. L.M performed and analyzed the time-resolved measurements with assistance from B.M. B.M wrote the original draft. B.M and A.D. wrote the Supplementary Materials. J.F conceptualized the idea. G.S., and J.F. acquired the funding, administrated, and supervised the project. All authors contributed to the interpretation of the results and the review and editing of the draft.

**Data availability**
The datasets generated and/or analyzed during the current study are available from the corresponding authors upon reasonable request.

**Additional information**
Reprints and permissions information is available online at www.nature.com/reprints. Correspondence and requests for materials should be addressed to B.M. (bmarzban@phys.ethz.ch) and J.F. (jfaist@ethz.ch).

**Competing financial interests**
The authors declare no competing financial interests.